\documentclass[final,leqno]{siamltexmm2}
\usepackage{amsmath,amssymb}

\usepackage{graphicx}
\usepackage{dcolumn}
\usepackage{bm}
\usepackage{epstopdf,overpic,color,rotating}
\usepackage{overpic}
\usepackage{gensymb}
\usepackage[table]{xcolor}
\usepackage[utf8]{inputenc}
\usepackage{textcomp,marvosym}

\usepackage{algorithm}
\usepackage{algpseudocode}
\algnewcommand\algorithmicinput{\textbf{Input:}}
\algnewcommand\Input{\item[\algorithmicinput]}

\newcommand*\samethanks[1][\value{footnote}]{\footnotemark[#1]}

\title{Model selection for dynamical systems via\\ sparse regression and information criteria}

\author{Niall M. Mangan$^\ddagger$\thanks{Department of Applied Mathematics, University of Washington, Seattle, WA. 98195.   \email{niallmm@uw.edu}.}
\and J. Nathan Kutz\samethanks[1] 
 \and Steven L. Brunton\thanks{Department of Mechanical Engineering, University of Washington, Seattle, WA. 98195. }
 \and Joshua L. Proctor\thanks{Institute for Disease Modeling, Bellevue, WA 98005, USA }}

\begin{document}
\maketitle
\newcommand{\slugmaster}{%
\slugger{siads}{xxxx}{xx}{x}{ }}

\begin{abstract}
We develop an algorithm for model selection which allows for the consideration of a combinatorially large
number of candidate models governing a dynamical system.  
The innovation circumvents a disadvantage of standard model selection which typically limits the number
candidate models considered due to the intractability of computing information criteria.
Using a recently developed sparse identification of nonlinear dynamics algorithm, the sub-selection of candidate models near the Pareto frontier allows for a tractable computation of AIC (Akaike information criteria) or BIC (Bayes information criteria) scores for the remaining candidate models.
The information criteria hierarchically ranks the most informative models, enabling the automatic and principled selection of the model with the strongest support in relation to the time series data. 
Specifically, we show that AIC scores place each candidate model in the {\em strong support}, {\em weak support} or {\em no support} category.  
The method correctly identifies several canonical dynamical systems,  including an SEIR 
(susceptible-exposed-infectious-recovered) disease model and the Lorenz equations, giving the correct
dynamical system as the only candidate model with strong support.
\end{abstract}

\section{Introduction}
\begin{figure}[t]
	\centering
	\includegraphics[width=\textwidth]{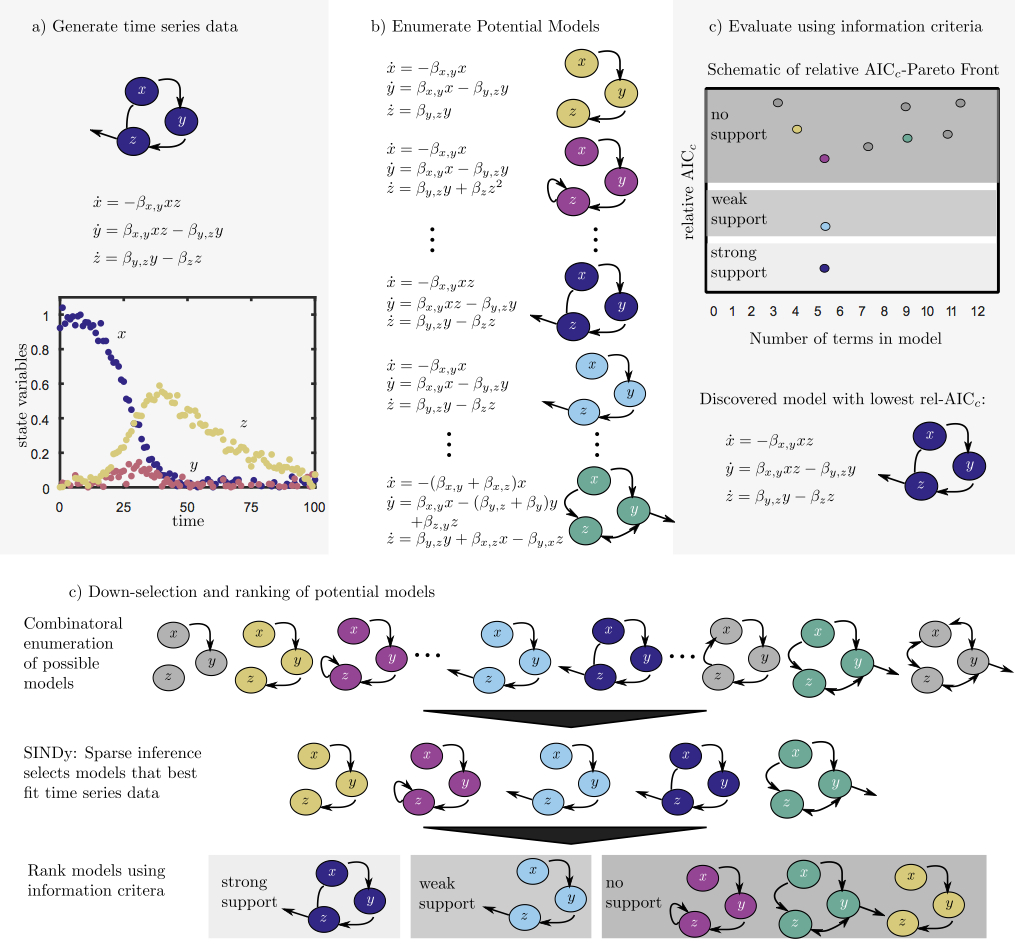}
	\caption{Schematic of model selection process, with a) data generation, b) generation of a set of potential models, and c) comparison fo the models as a function of the number of terms in the model and relative Akaike information criteria (AIC$_c$). Section c) shows how models are down-selected from a combinatorially large model space using sparse identification of nonlinear dynamics (SINDy) and then further sub-selected and ranked using information criteria.}
	\label{fig:overview}
\end{figure}

Nonlinear dynamical systems theory has provided a fundamental characterization and understanding of phenomenon across the physical, engineering  
and biological sciences.  Traditionally, simplified models
are posited by domain experts, and simulations and analysis are used to explore the underlying dynamical behavior which may
include chaotic dynamics (e.g. Lorenz equations), nonlinear oscillations (e.g. van der Pol, Duffing), and/or bifurcations.  The emergence of data-driven modeling methods provides an alternative framework for the discovery and/or inference of governing nonlinear dynamical equations.  From this perspective, governing models are posited from time-series measurement data alone.  
The recent \emph{sparse identification of nonlinear dynamics} (SINDy) method~\cite{Brunton2016pnas} uses sparse regression and a Pareto analysis to correctly discover parsimonious governing equations from a combinatorially large set of potential dynamical models.  
This methodology can be generalized to spatio-temporal systems~\cite{Rudy2016arxiv,schaeffer2017} and dynamical systems characterized with rational function nonlinearities which often occur in biological networks~\cite{Mangan2016ieee}.    Although previously suggested~\cite{Mangan2016ieee}, no explicit connection between the sparse selection process and information theoretic criteria has been established.  Information criteria is the standard statistical method established for the model selection process.  
In this manuscript, we demonstrate that the Akaike information criteria (AIC) and/or Bayesian information criteria (BIC) can be connected with the SINDy architecture to hierarchically rank models on the Pareto front for automatic selection of the most informative model.  
As outlined in Fig.~\ref{fig:overview}, the AIC/BIC metrics can be used to correctly infer dynamical systems for a given time-series data set from a combinatorially large set of models. To our knowledge, this is the first explicit demonstration of how information theory can be exploited for the identification of dynamical systems.

Model selection is a well established statistical framework for selecting a model from a set of candidate models given time series data, with information theory providing a rigorous criteria for such a selection process.
As early as the 1950s, a measure of information loss between empirically collected data and model generated data was proposed to be computed using the Kullback-Leibler (KL) divergence \cite{kullback1951,kullback1959}.  Akaike built upon this notion to establish a {\em relative} estimate of information loss across models that balances model complexity, and goodness-of-fit~\cite{akaike1973,akaike1974}.  This allowed for a principled model selection criteria through the Akaike information criteria (AIC).  The AIC was later modified by G. Schwarz to define the more commonly used Bayes information criteria (BIC)~\cite{schwarz1978estimating}.  Both AIC and BIC compute the maximum log likelihood of the model and impose a penalty:  
AIC adds the number of free parameters $k$ of the posited model, while BIC adds half of $k$ multiplied by the log of the number of data points $m$.  Much of the popularity of BIC stems from the fact that it can be rigorously proved to be a consistent metric~\cite{schwarz1978estimating}.  Thus if a number of models $q$ are proposed, with one of them being the true model, then as $m\rightarrow \infty$ the true model is selected as the correct model with probability approaching unity.  Regardless of the selection criterion, AIC or BIC, they both provide a relative estimate of information loss across a selection of $m$ models, balancing model complexity and goodness-of-fit~\cite{burnham2002}.  

Although successful and statistically rigorous, model selection in its standard implementation is typically performed on $q$ predetermined candidate models, where $q$ is often ten or less\cite{burnham2002,Kuepfer2007,Hjorth2008,Schaber2011,woodward2004epidemiology,Blake22072014}.  For modern applications to dynamical systems where rich, high-fidelity time series data can be acquired, the restriction on the number of models limits the potential impact of AIC/BIC metrics for discovering the correct nonlinear dynamics.  Instead, it is desired to consider a combinatorially large set of potential dynamical models as candidates, thus enforcing that $q\gg 1$.  
This is computationally intractable with standard model selection, as each of the models from  the combinatorially large set would have to be simulated and then evaluated for a BIC/AIC score.  

As an alternative, sparse regression techniques, embodied by the {\em lasso} (least absolute shrinkage and selection operator) method of Tibshirani~\cite{Tibshirani:1996}, have enabled variable selection algorithms capable of optimally choosing among a combinatorially large set of potential predictors.  Specifically, a lasso regression analysis, or one of its many generalizations and variants, performs both variable selection and regularization in order to enhance the prediction accuracy and interpretability of the statistical model it produces.  Such a mathematical tool provides a critically enabling framework for model selection, in particular for identifying dynamical systems.

In this work we demonstrate a new mathematical framework that leverages information criteria for model selection with sparse regression for evaluating a combinatorially large set of candidate models.
Specifically, we circumvent a direct computation of information criteria for the combinatorially large set
of models by first sub-selecting candidate functional forms 
from which are most consistent with the time series data. 
 Thus we integrate two maturing fields of statistical analysis:  (i) sparse regression for nonlinear systems identification via SINDy and (ii) model selection via information criteria.   Our algorithm is demonstrated to produce a robust procedure for discovering parsimonious, nonlinear dynamical systems from time series measurement data alone.   We demonstrate the methodology on a number of important examples, including the SEIR 
(susceptible-exposed-infectious-recovered) disease model and the Lorenz equations, and demonstrate its efficacy as a function of noise, length of time series and other key regression factors.  Our sparse selection of dynamical models from information theory criteria ranks the candidate models and further shows that the correct model is strongly supported by the AIC/BIC metrics.   Ultimately, the method provides a cross-validated and ranked set of candidate nonlinear dynamical models for a given time-series of measurement data, thus enabling data-driven discovery of the underlying governing equations.


\section{Background}

\subsection{Model selection via information criteria}
The process of model selection fundamentally enables the connection of observations or {\em data} to a mathematical model.   
Further, a well-selected model, which describes a governing law or physical principle underlying the system process, can be utilized for prediction outside of the sampled data and parameter configuration~\cite{burnham2002}.  
The substantial challenge facing the selection process is discovering the {\em best} predictive model from a combinatorially large space of available models.  
To emphasize the enormity of this task, consider the number of possible polynomial models up to degree $4$ with $5$ state variables.  Approximately $10^{38}$ models would need to be constructed, fit to the data, and compared according to a goodness-of-fit metric~\cite{Mangan2016ieee}.  
Thus, model selection quickly becomes computationally intractable for a modest number of variables and polynomial degree.

Typically, a sub-selection of models occurs based on prior scientific knowledge of the process to produce a subset, $\mathcal{O}(10)$, of heuristically defined {\em candidate models}~\cite{burnham2002,Kuepfer2007,Hjorth2008,Schaber2011,woodward2004epidemiology,Blake22072014}.  
Recent research has focused on automatically expanding the number of candidate models~\cite{Schmidt2009science,Buchel2013,Cohen2015}.  
Once a subset of models is chosen, the model selection procedure balances the goodness-of-fit with model complexity, i.e. the number of free parameters. 
A wide-variety of rigorous statistical metrics have been developed to balance model parsimony and predictive power including popular methods such as the Akaike information criterion (AIC)~\cite{akaike1973,akaike1974}, Bayesian information criterion (BIC)~\cite{schwarz1978estimating}, cross-validation (CV)~\cite{bishop2006pattern}, deviance information criterion (DIC)~\cite{linde2005dic}, and minimum description length (MDL)~\cite{rissanen1978modeling}.  
Methods such as AIC explicitly balance parsimony and relative information loss across models, penalizing the number of parameters in the model to avoid overfitting.  

In this manuscript, we utilize the ubiquitous and well-known AIC as the statistical criterion for comparing candidate models.  
The AIC value for each candidate model $j$ is:
\begin{equation}
AIC_j = 2 k  - 2 \ln ({L}(\mathbf{x},\hat{\mu})),
\label{Eq.AIC}
\end{equation}
where $L(\mathbf{x},{\mu})=P(\mathbf{x}|\mu)$ is the likelihood function (conditional probability) of the observations $\mathbf{x}$ given the parameters $\mu$ of a candidate model, $k$ is the number of free parameters to be estimated, and $\hat{ \mu}$ is the best-fit parameter values for the data~\cite{akaike1973,akaike1974}.  
In practice, the AIC requires a correction for finite sample sizes given by
\begin{equation}
AIC_c = AIC + 2(k+1)(k+2)/(m-k-2),
\label{Eq.AIC_rel}
\end{equation}
where $m$ is the number of observations.  
A common likelihood function uses the residual sum of squares (RSS), given by $RSS = \sum_{i=1}^m (y_i- g(x_i;\mu))^2$, where $y_i$ are the observed outcomes, $x_i$ are the observed independent variables, and $g$ is the candidate model.  The RSS is a well-known objective function for least squares fitting. In this case AIC can be expressed as $AIC = m \ln(RSS/m) +2k$ ~\cite{burnham2002}.   
Note that \eqref{Eq.AIC} penalizes, by increasing the AIC score, the models that have a large number of free parameters and which are unable to capture the characteristics of the observed data.

\subsection{SINDy and sparse model selection}
Identifying dynamical systems models from data is increasingly possible with access to high-fidelity data from simulations and experiments.  
With traditional methods, only a small handful of model structures may be posited and fit to data via regression.  
Indeed, simultaneous identification of both the structure and parameters of a model generally requires an intractable search through combinatorially many candidate models.  
Genetic programming has been recently used to determine the structure and parameters of dynamical systems~\cite{Bongard2007pnas,Schmidt2009science,Quade2016arxiv} and control laws~\cite{Duriez2016book}, enabling the efficient search of complex function spaces.  
Sparsity-promoting techniques have also been employed to simultaneously identify the structure and parameters of a dynamical system model.  
Compressed sensing was first used to determine the active terms in the dynamics~\cite{Wang2011prl}, although it does not work well with overdetermined systems that arise when measurements are abundant.  
In contrast, the sparse identification of nonlinear dynamics (SINDy) algorithm~\cite{Brunton2016pnas} uses sparsity-promoting regression, such as the {\em lasso}~\cite{Tibshirani:1996} or sequential thresholded least-squares algorithm~\cite{Brunton2016pnas}, to identify nonlinear dynamical systems from data in overdetermined situations.  

Here we review the SINDy architecture for identifying nonlinear dynamics from data.  
The general observation underlying SINDy is that most dynamical systems of a state $\mathbf{x}\in\mathbb{R}^n$, 
\begin{equation}
\frac{d}{dt}\mathbf{x}(t) = \mathbf{f}(\mathbf{x}(t)), \label{Eq:Dynamics}
\end{equation}
have only a few active terms in the dynamics, making them \emph{sparse} in a suitable function space.  
To identify the structure and parameters of the model, a set of candidate symbolic functions are first concatenated into a  library $\boldsymbol{\Theta}(\mathbf{x}) = \begin{bmatrix} \theta_1(\mathbf{x}) & \cdots & \theta_p(\mathbf{x})\end{bmatrix}$.  With time-series data ${\bf X}\in\mathbb{R}^{m\times n}$, where each row is a measurement of the state $\mathbf{x}^T(t_k)$ in time, it is possible to evaluate the candidate function library $\boldsymbol{\Theta}(\mathbf{X})\in\mathbb{R}^{m\times p}$ at the $m$ time points.  Finally, with derivative data $\dot{\bf {X}}\in\mathbb{R}^{m\times n}$, either measured or obtained by numeric differentiation, it is possible to pose a regression problem:
\begin{equation}
\frac{d}{dt}\mathbf{X} = \boldsymbol{\Theta}(\mathbf{X})\boldsymbol{\Xi}.
\end{equation}
The few active terms in the dynamics, given by the nonzero entries in the columns of $\boldsymbol{\Xi}$, may be identified using sparse regression.  
In particular, the sparsest matrix of coefficients $\boldsymbol{\Xi}$ is determined that also provides a good model fit, so that $\|\dot{\bf {X}}-\boldsymbol{\Theta}({\bf X})\boldsymbol{\Xi}\|_2$ is small.  
Sparse regression has the added benefit of avoiding overfitting, promoting stability and robustness to noise.  
Since the original SINDy method, there have been numerous innovations and extensions to handle rational function nonlinearities~\cite{Mangan2016ieee}, partial differential equations~\cite{Rudy2016arxiv}, highly corrupted data~\cite{Tran2016arxiv}, and to build Galerkin regression models in fluids~\cite{Loiseau2016arxiv}.  The method is also connected to the dynamic mode decomposition (DMD)~\cite{Kutz2016book} if only linear functions are used in $\boldsymbol{\Theta}$.  

In most sparse regression algorithms, there is a parameter that determines how aggressively sparsity is promoted.  
The successful identification of the model in Eq.~\eqref{Eq:Dynamics} hinges on finding a suitable value of this sparsity-promoting parameter.  
Generally, a the parameter value is swept through, and a Pareto front is used to select the most parsimonious model.  
However, the Pareto frontier may not have a sharp elbow or may instead have a cluster of models near the elbow, thus compromising the automatic nature of the model selection using SINDy alone.

\section{Methods}

Our algorithm integrates sparse regression for nonlinear system identification with model selection via information criteria.  
This approach enables the automatic identification of a single best-fit model from a combinatorially large model space.  
In the first step of the algorithm, the SINDy method provides an initial sub-selection of models from a combinatorially large number of candidates.
The sub-selection of candidate models {\em near} the Pareto frontier is critically enabling as it is computationally intractable to simulate and compare against the time series data.
Importantly, the sub-selection can take the number of candidate modes from $10^{9}$ (for our 2-D cubic example) to a manageable $10^{2}$.  
This then allows for a tractable computation of AIC or BIC scores for the remaining candidate modes.
The information criteria hierarchically ranks the most informative models, enabling the automatic selection of the model with the strongest support. 
This is in contrast to a standard Pareto front analysis which looks for a parsimonious model at the elbow of the error versus complexity curve.
However, the Pareto frontier may not have a sharp elbow or may instead have a cluster of models near the elbow.
Algorithm~\ref{alg}, using AIC as our information criteria, is executed for model selection.  Figure~\ref{fig:overview} illustrates the algorithm in practice.

    \begin{algorithm}[t]
		\caption{SINDy -- AIC}
		\label{cSVDalgorithm}
		\begin{algorithmic}[1]
			\Input Data matrix and time derivative:  $\mathbf{X}, \dot{\mathbf{X}} \in \mathbb{R}^{m \times n}$ 
			\Procedure{SINDy--AIC}{$\mathbf{X}$, $\dot{\mathbf{X}}$}
			\State $\boldsymbol{\Theta}\in \mathbb{R}^{m \times p}$  $\gets$ library($\mathbf{X}$)
			\Comment{Generate library that contains $p$ candidate terms.}
	\For{$\lambda (j) \in \{ \lambda_0, \lambda_1, \cdots, \lambda_q \}$}
	\Comment{Search over sparsification parameter $\lambda$.}
			\State Model($j$) $\gets$ SINDy($\dot{\mathbf{X}}$, $\boldsymbol{\Theta}$, $\lambda(j)$ )
			\Comment{Identify sparse terms in $\boldsymbol{\Theta}$ for model.}
			\State $\mathbf{X}'$ $\gets$ simulate (Model($j$))
			\Comment{Numerically integrate dynamical system (expensive).}
			\State IC($j$) $\gets$  AIC( $\mathbf{X}, \mathbf{X}'$)
			\Comment{Compute Akaike information criteria.}
			\EndFor
			\State [inds,vals] $\gets$ sort(IC)
			\Comment{Rank models by AIC score.}
			\State \textbf{return} Model(inds(1))
			\Comment{Return best-fit model.}
			\EndProcedure
		\end{algorithmic}
		\label{alg}
\end{algorithm}	

When evaluating dynamical systems models, there is some ambiguity about what constitutes an "observation." We take time series data for a given set of initial conditions to be an observation, rather than taking each measurement at each time point. To obtain a representative error for the $i$th time-series observation, we calculate the average absolute error over the entire time series: $E_{\mbox{avg}} = \sum_{\tau}|y_i^\tau - g(x_i^\tau;\mu)|$. We then substitute this representative error in for $(y_i- g(x_i;\mu)$, and take the sum of the squares so that  $AIC = m \:  \ln ((\sum_{i=1}^m E_{avg}(x_i,y_i))/m) +2k$. We use this value for AIC in (\ref{Eq.AIC_rel}).

The candidate models with the lowest scores are ranked as the most likely. 
To be more precise, the AIC scores for each candidate model can have a wide range of values which require a rescaling by the minimum AIC value ($AIC_{min}$)~\cite{burnham2002,burnham2004}. 
The rescaled AIC values $\Delta_j = AIC_i - AIC_{min}$ can be directly interpreted as a strength-of-evidence comparison across models.  
Models with $\Delta_j \leq 2$ have {\em strong support}, $4 \leq \Delta_j \leq 7$ have {\em weak support}, and $\Delta_j \geq 10$ have {\em no support}~\cite{burnham2004}.  These rankings allow for a principled procedure for retaining or rejecting models within the candidate pool of models.
For time series data with enough data samples, large enough signal to noise and/or sufficiently large set of candidate models, only one model is typically strongly supported by AIC metrics.  In the examples used to demonstrate the method, the only strongly supported candidate model is, in fact, the correct model.

\section{Results: Model selection}
\subsection{1-D polynomial model}
\begin{figure}[t]
	\centering
	\includegraphics[width=\textwidth]{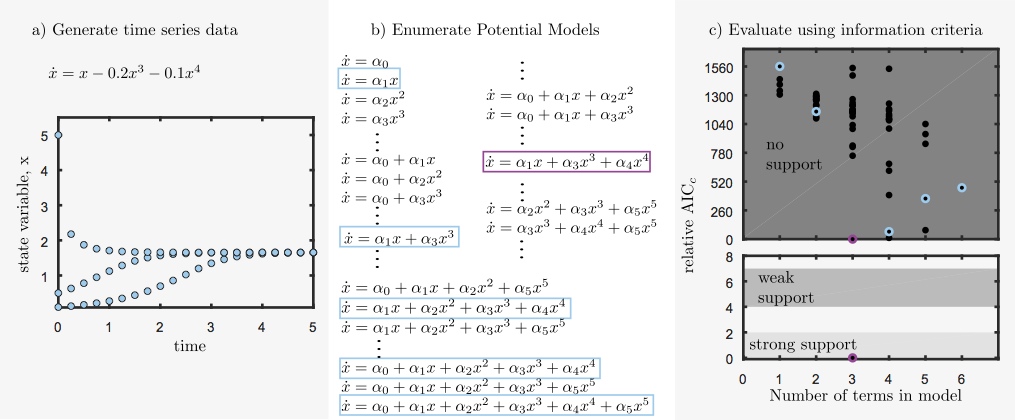}
	\vspace*{-.1in}
	\caption{Selection of model for single variable, $x$, polynomial system. a) 3 computationally generated time series with additive noise $\epsilon = 0.001$. b) Combinatorial model possibilities with with those selected by SINDy highlighted in blue. c) Relative AIC$_c$ criteria for all possible models (black dots), and those found by SINDy (blue circles). Lower plot magnifies strong and weakly supported AIC$_c$ range, containing only correct model (magenta circle).}
	\label{fig:1Dpoly}
\end{figure}

We demonstrate the relationship between SINDy model selection and AIC ranking procedure, as illustrated in Fig.~\ref{fig:overview}~d, with a simple single state variable model with three polynomial terms,
\begin{equation}
\dot{x} = x - 0.2 x^3 - 0.1 x^4.
\end{equation}
We compute three time-series for this model, as shown in Fig.~\ref{fig:1Dpoly}~a.

Assuming a feature library with up to 5th order polynomials, or $d =6$, there are $\sum_{i=1}^{6} \binom{6}{i} = 63$ possible models. A representation of the combinatorial set of models is shown in Fig.~\ref{fig:1Dpoly}~b, and we perform least-squares fitting for the coefficients of each model. All models are cross-validated using 100 initial conditions, and relative AIC$_c$ scores are calculated. This combinatorially complete set of models (black dots) populate the AIC$_c$ Pareto front in Fig.~\ref{fig:1Dpoly}~c. The relative AIC$_c$ metric successfully characterizes the true, 3 term model as the only model with "strong support." All other models within the full combinatorial space fall within the "no support" range.

 SINDy enables us to sub-select a set of models from the feature library and, for this simple system, we can compare against the combinatorial set. The models selected by SINDy are highlighted in blue in Fig.~\ref{fig:1Dpoly}~c and d. SINDy finds the correct 3 term model, as well as models with 1, 2, 4, 5, and 6 terms each. These models and coefficients exactly match a subset of those found by combinatorial least-squares, which is expected given that this implementation of SINDy uses least-squares to fit the coefficients. Notably, the 1 and 2-term models selected by SINDy are the most meaningful reductions of the true underlying system (with smaller coefficients set to zero), rather than the 1 and 2 term models with the lowest error in the full feature space.

\subsection{2-D cubic model}
\begin{figure}[t]
\hspace*{-.7in}
	\includegraphics[width=1.3\textwidth]{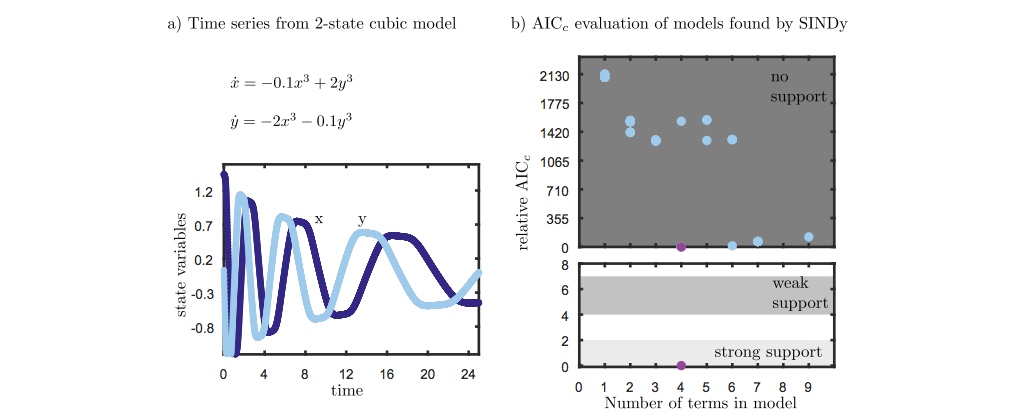}
	\vspace*{-.2in}
	\caption{Evaluation of SINDy selected models for 2-D cubic system. a) computationally generated time series from single set of initial conditions with additive noise $\epsilon = 0.001$. b) Relative AIC$_c$ criteria for models found by SINDy (blue circles). Magnification in lower plot shows strong and weakly supported AIC$_c$ range contains only the correct model (magenta circle)..}
	\label{fig:2Dcubic}
\end{figure}

 For larger systems, enumerating all possible models represented in a given feature library would be computationally infeasible, but by using SINDy a sub-selection of the most relevant models are generated and selected. For example, consider the relatively simple example of a 2-state variable system ($n=2$) with a 5th order polynomial library ($d=6$). In this case there are $N_m = {n + d \choose d} = 28$, possible monomials, and $N_p = \sum_{i=1}^{N_m} { N_m \choose i}  = 268,435,455$ potential models. Figure~\ref{fig:2Dcubic} demonstrates the results of performing SINDy and AIC evaluation on a cubic model.

With only a single time-series for each state variable (x and y) as input (Fig.~\ref{fig:2Dcubic}~a), the SINDy selected models with varying number of terms (blue circles) include the correct model (magenta circle). Using 100 randomly selected initial conditions for cross-validation, relative AIC$_c$ to ranks the correct model as strongly supported, and all other models as having no support. 

\subsection{3-D Disease transmission model}
\begin{figure}[t]
\hspace*{-.7in}
	\includegraphics[width=1.3\textwidth]{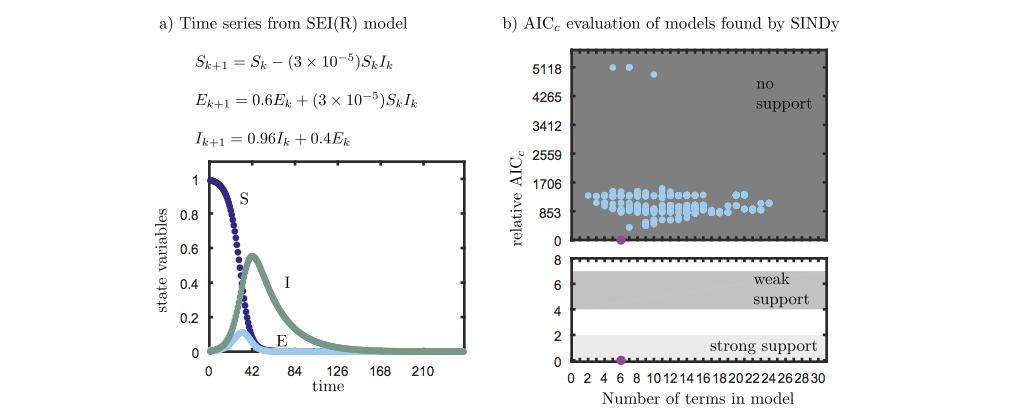}
	\vspace*{-.2in}
	\caption{Evaluation of SINDy selected models for 3 state variable SEIR model. a) computationally generated time series with additive noise $\epsilon = 2.5\times 10^{-4}$. b) Relative AIC$_c$ criteria for models found by SINDy. Magnification in lower plot shows strong and weakly supported AIC$_c$ range contains only the correct model (magenta circle).}
	\label{fig:SEIR}
\end{figure}

Next we apply our method to the susceptible-exposed-infectious-recovered (SEIR) disease transmission model. Models of this type are often used to determine disease transmission rates, detect outbreaks and develop intervention strategies. However, generating the appropriate model for interactions between different populations is currently done heuristically and then evaluated using information criteria \cite{Blake22072014}.  Using SINDy with AIC evaluation would provide {\em data-driven} model generation and selection from a wider library of possible interaction terms. As a first step, we apply SINDy with AIC to a discrete, deterministic SEIR model as shown in Fig.~\ref{fig:SEIR}.

We input a single time series representing an outbreak for the $S$, $E$, and $I$ state variables (n =3), and provide a library of polynomial terms up to 2nd order (d= 3). For this example, the total number of models represented in the library is $N_p = \sum_{i=1}^{N_m} { N_m \choose i}  = 1023$ with $N_m = {n + d \choose d} = 10$ possible monomials. A complication of the SEIR system is that $R$ is a redundant state variable; $S$, $E$, and $I$ have no dependence on $R$, and $R$ depends only on a term already represented in the $I_{k+1}$ equation $R_{k+1} = R_k + 0.04 I_k$. A result of this redundancy, SINDy cannot find the correct equations with $R$ included in the library. Without $R$, SINDy selects a set of models from 1023 possible in the library (blue circles), and with 100 cross-validation measurements the relative-AIC$_c$ evaluation ranks only the correct 6-term model as having any support (magenta circle) in Fig.~\ref{fig:SEIR}~b.


\subsection{Lorenz model}
\begin{figure}[t]
	\hspace*{-.7in}
	\includegraphics[width=1.3\textwidth]{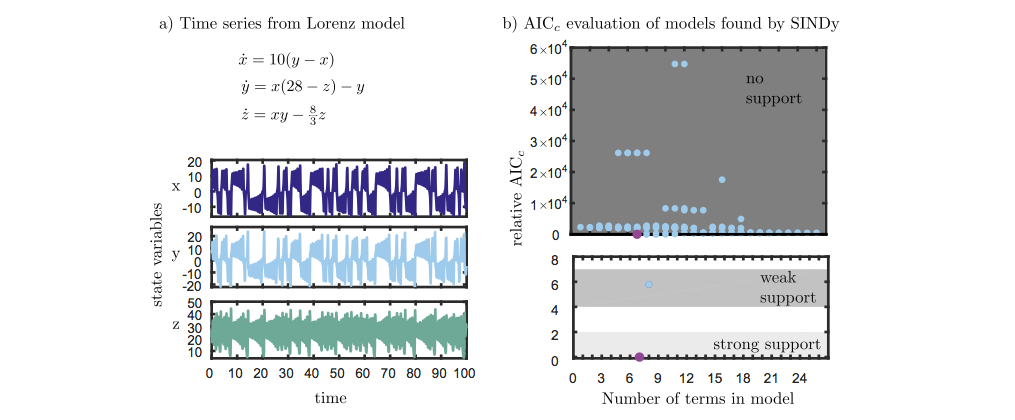}
	\vspace*{-.2in}
	\caption{Evaluation of SINDy selected models for 3 state variable Lorenz model. a) computationally generated time series with additive noise $\epsilon = 0.001$. b) Relative AIC$_c$ criteria for models found by SINDy. Magnification in lower plot shows strong and weakly supported AIC$_c$ range. The correct model lies in the strong support range and a model with an additional small term in the weak support range.}
	\label{fig:Lorenz}
\end{figure}

As a final example, we demonstrate SINDy with AIC on the chaotic Lorenz system \cite{Lorenz1963}. Using a library of polynomials up to 2nd order (d=3) for the 3-state variable system (n = 3), there are once again $N_p = 1023$ models represented in the function library. Providing one time-series for each state variable, as shown in Fig.~\ref{fig:Lorenz}~a, SINDy recovers a subset of these models (circles in Fig.~\ref{fig:Lorenz}~b). 
Using 100 cross-validation measurements on each model, the relative-AIC$_c$ criteria ranks the correct 7-term model as having strong support (magenta circle). Unlike in previous examples, one other model is ranked as having "weak support" (blue circle). This model has an additional small constant term in the equation for $x$: $\dot{x} = 8.5\times 10^{-6} + 10(y-x)$. 

A possible explanation for the level of support for this model is the chaotic nature of the Lorenz system. Even when the recovered model is correct, small variations in the recovered coefficients ($\approx 10^{-6}$ for this case), will cause the calculated time-series for the recovered model to diverge from the "true" model after some length of time ($>1$ unit time for these parameters). For the example in Fig.~\ref{fig:Lorenz}~b, the cross-validation uses time series of length $t=5$ (arb. units). In a true model-selection situation, we would not know this characteristic length scale ahead of time, and a sensitivity analysis would need to be preformed. We discuss this and other challenges to practical implementation in the next section.

\section{Practical implementation: noise and number of measurements}
\begin{figure}[t]
	\hspace*{-.7in}
	\includegraphics[width=1.2\textwidth]{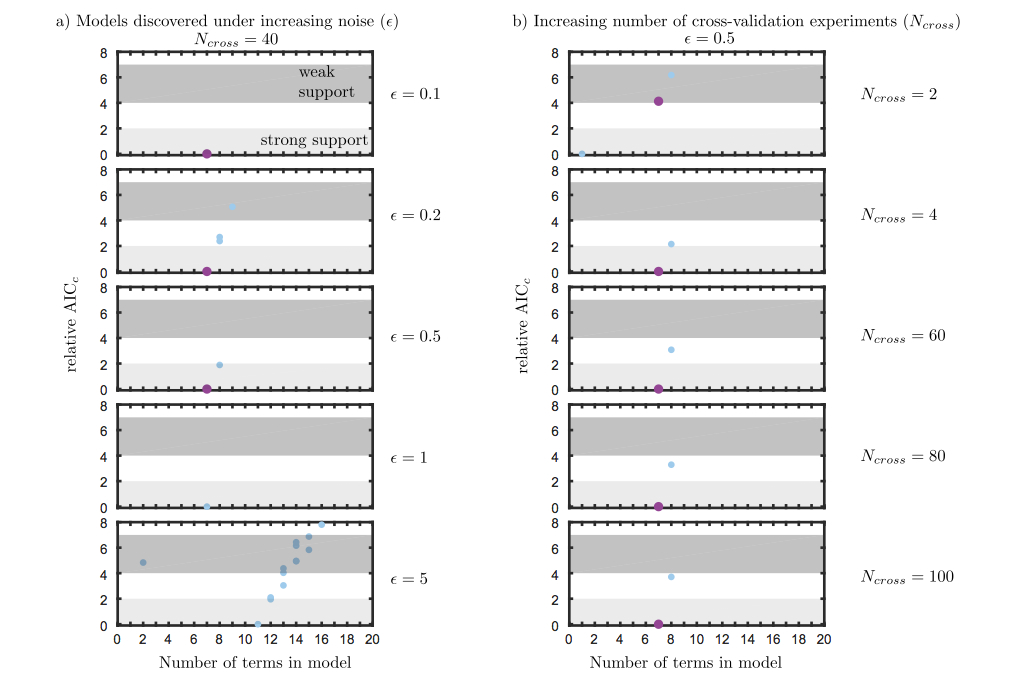}
	\vspace*{-.2in}
	\caption{Evaluation of SINDy selected models for 3 state variable Lorenz model under varying a) noise conditions and using b) increasing number of cross-validation experiments to calculate AIC$_c$. Strong and weakly supported relative-AIC$_c$ range is shown.}
	\label{fig:Noise}
\end{figure}

SINDy with AIC ranking can successfully select the correct model for a variety of known systems, given low enough measurement noise and a large amount of data for cross-validation. Under practical conditions, the signal-to-noise may be lower than desired and the amount of data available for cross-validation may be restricted. In Fig.~\ref{fig:Noise} we show the effects of increasing noise and number of cross-validation experiments on the selection of the correct model for the Lorenz system. Reading from the top of Fig.~\ref{fig:Noise}~a, for low noise, $\epsilon =0.1$, only the correct model (magenta) falls within the supported (strong or weak) range. Increasing the noise to $\epsilon = 0.2$ and $0.5$ cause other models to descend into the weak support regime and eventually into the strong support range. Around this level of noise, the relative AIC$_c$ scores for the incorrect models are very sensitive to the random additive noise. Repeating the computation for different instances of randomly generated measurement noise causes the position of these models to fluctuate (data not shown), although the true model maintains the lowest score (over 10 instances). This suggests that data sub-sampling could be used to test for models with noise-fitted terms. 

Above a certain level of noise ($\epsilon = 1$) the method is unable to robustly select the correct model, and for even higher levels of noise ($\epsilon = 5$) a larger number of incorrect models appear to have support. For the particular case of the Lorenz system used here, SINDy is unable to select the correct model, and therefore the true model is not evaluated by AIC$_c$ at these noise levels. As we are using a relative AIC$_c$ score, the score of the lowest model will always be zero, even when the actual error between that model and the data is very high. These examples highlight the importance of examining the error between the model-generated time-series against and the data in addition to the relative AIC$_c$ metric.

The number of time-series used for cross-validation also has an impact on the relative AIC$_c$ score between the true model and other models given strong support as shown in Fig.~\ref{fig:Noise}~b. For $\epsilon = 0.5$ and only 2 time-series used for cross-validation and AIC$_c$ calculation, the correct model is selected by SINDy, but a simpler (and incorrect) 1-term model is assigned a lower score. Given only 2 more time-series for cross-validation, ($N_{cross} =4$), the correct model has the lowest score. Increasing the number of cross-validation measurements used for evaluation to $N_{cross} = 60, 80,$ and $100$ increases the relative score of the incorrect 8-term model to the true 7-term model. 

The success of the method also relies on the duration and sampling of the time series, especially in the case of chaotic systems like the Lorenz model. Here the cross-validation experiments run from $t=0$ to $t_{end}=1$ (arb. units), as opposed to those in Fig.~\ref{fig:Lorenz} which ran to  $t_{end} = 5$. Even with higher noise ($\epsilon = 0.1$ compared to $\epsilon = 0.001$), only the true model is supported. Again, sensitivity to sub-sampling of data can help differentiate between noise-fitting and mechanistically essential terms.

\section{Discussion and conclusions}

The integration of mathematical techniques advocated here,  (i) sparse regression for nonlinear systems identification via SINDy and (ii) model selection via information criteria, provides a new paradigm for model selection of dynamical systems. Algorithmically, the critical methods combine as follows.  In the first step of the algorithm, the SINDy method, which is based upon sparse regression, provides an initial sub-selection of models from a combinatorially large number of candidate models.   
The selection of candidate models is critically enabling 
as it reduces the number of potential models to a manageable number, which can each be evaluated through simulation and comparison to the time series data.  Indeed, the remaining candidate models, which are now on the order of ten models, are each evaluated using information criteria such as AIC or BIC.

The candidate models with the lowest scores are ranked as the most likely. Specifically, in what follows, we work with the AIC score and show that these scores place each candidate model in the {\em strong support}, {\em weak support} or {\em no support} category.  For time series data with enough data samples, large enough signal to noise and/or a sufficiently large set of candidate models, only one model is typically strongly supported by AIC metrics.  In the examples used to demonstrate the method, the only strongly supported candidate model is, in fact, the correct model.  

The method presented provides an important contribution to standard model selection as well as to the SINDy paradigm.  In particular, each of these methods has a significant shortcoming.  In model selection, the shortcoming is centered around the inability of the standard AIC/BIC criteria to assess a combinatorially large set of candidate models.  For SINDy, the sparse selection process for identifying the underlying dynamical systems lacks a principled method for selecting the correct dynamical model.  The algorithm here circumvents both of these shortcomings.  Specifically, the sparse regression of SINDy allows for the consideration of a combinatorially large number of candidate models.  The sub-selected set of models can then each be evaluated using information criteria to select the correct dynamical system.  
The connection between information criteria and automatic model selection can also be integrated with genetic algorithms for selecting the structure and parameters of dynamical systems~\cite{Bongard2007pnas,Schmidt2009science,Quade2016arxiv,Duriez2016book}. 
The process can be automated for data-driven discovery of physical principles and laws of motion, which is now often referred to as the 4th paradigm of science~\cite{gray}.

\section*{Acknowledgements} JNK acknowledges support from the Air Force Office of Scientific Research (FA9550-15-1-0385).  SLB and JNK acknowledge support from the Defense Advanced Research Projects Agency (DARPA contract HR0011-16-C-0016).  JLP and NMM would like to thank Bill and Melinda Gates for their active support of the Institute for Disease Modeling and their sponsorship through the Global Good Fund.

\bibliographystyle{prsb}
\bibliography{references,biball}

\begin{thebibliography}{10}
\expandafter\ifx\csname urlstyle\endcsname\relax
  \providecommand{\doi}[1]{doi:\discretionary{}{}{}#1}\else
  \providecommand{\doi}{doi:\discretionary{}{}{}\begingroup
  \urlstyle{rm}\Url}\fi

\bibitem{Brunton2016pnas}
Brunton, S.~L., Proctor, J.~L. \& Kutz, J.~N., 2016 Discovering governing
  equations from data by sparse identification of nonlinear dynamical systems.
\newblock \emph{Proceedings of the National Academy of Sciences} \textbf{113},
  3932--3937.

\bibitem{Rudy2016arxiv}
Rudy, S.~H., Brunton, S.~L., Proctor, J.~L. \& Kutz, J.~N., 2016 Data-driven
  discovery of partial differential equations.
\newblock \emph{arXiv preprint arXiv:1609.06401} .

\bibitem{schaeffer2017}
Schaeffer, H., 2017 Learning {PDE} via data discovery and sparse optimisation.
\newblock \emph{Proc. Roy. Soc. A (to appear)} .

\bibitem{Mangan2016ieee}
Mangan, N.~M., Brunton, S.~L., Proctor, J.~L. \& Kutz, J.~N., 2016 Inferring
  biological networks by sparse identification of nonlinear dynamics.
\newblock \emph{To appear in the IEEE Transactions on Molecular, Biological,
  and Multi-Scale Communications} .

\bibitem{kullback1951}
Kullback, S. \& Leibler, R.~A., 1951 {On Information and Sufficiency}.
\newblock \emph{The Annals of Mathematical Statistics} \textbf{22}, 79--86.
\newblock ISSN 0003-4851.
\newblock (\doi{10.1214/aoms/1177729694}).

\bibitem{kullback1959}
Kullback, S., 1959 \emph{{Information Theory and Statistics}}.

\bibitem{akaike1973}
Akaike, H., 1973 {Information theory and an extension of the maximum likelihood
  principle}.
\newblock In \emph{Petrov, B.N.; Cs{\'{a}}ki, F., 2nd International Symposium
  on Information Theory, Tsahkadsor, Armenia, USSR, September 2-8, 1971,}, pp.
  267--281. Budapest: Akad{\'{e}}miai Kiad{\'{o}}.

\bibitem{akaike1974}
Akaike, H., 1974 {A New Look at the Statistical Model Identification}.
\newblock \emph{IEEE Transactions on Automatic Control} \textbf{19}, 716--723.
\newblock ISSN 15582523.
\newblock (\doi{10.1109/TAC.1974.1100705}).

\bibitem{schwarz1978estimating}
Schwarz, G., 1978 Estimating the dimension of a model.
\newblock \emph{The annals of statistics} \textbf{6}, 461--464.

\bibitem{burnham2002}
Burnham, K. \& Anderson, D., 2002 \emph{{Model Selection and Multi-Model
  Inference}}.
\newblock Springer, 2nd edition.

\bibitem{Kuepfer2007}
Kuepfer, L., Peter, M., Sauer, U. \& Stelling, J., 2007 {Ensemble modeling for
  analysis of cell signaling dynamics.}
\newblock \emph{Nature biotechnology} \textbf{25}, 1001--1006.
\newblock ISSN 1087-0156.
\newblock (\doi{10.1038/nbt1330}).

\bibitem{Hjorth2008}
Claeskens, G. \& Hjorth, N.~L., 2008 \emph{{Model Selection and Model
  Averaging}}.
\newblock Cambridge University Press.

\bibitem{Schaber2011}
Schaber, J., Fl{\"{o}}ttmann, M., Li, J., Tiger, C.~F., Hohmann, S. \& Klipp,
  E., 2011 {Automated ensemble modeling with modelMaGe: Analyzing feedback
  mechanisms in the Sho1 branch of the HOG pathway}.
\newblock \emph{PLoS ONE} \textbf{6}, 1--7.
\newblock ISSN 19326203.
\newblock (\doi{10.1371/journal.pone.0014791}).

\bibitem{woodward2004epidemiology}
Woodward, M., 2004 \emph{{Epidemiology: Study Design and Data Analysis, Second
  Edition}}.
\newblock Chapman {\&} Hall/CRC Texts in Statistical Science. Taylor {\&}
  Francis.
\newblock ISBN 9781584884156.

\bibitem{Blake22072014}
Blake, I.~M., Martin, R., Goel, A., Khetsuriani, N., Everts, J., Wolff, C.,
  Wassilak, S., Aylward, R.~B. \& Grassly, N.~C., 2014 {The role of older
  children and adults in wild poliovirus transmission}.
\newblock \emph{Proceedings of the National Academy of Sciences} \textbf{111},
  10604--10609.
\newblock ISSN 0027-8424, 1091-6490.
\newblock (\doi{10.1073/pnas.1323688111}).

\bibitem{Tibshirani:1996}
Tibshirani, R., 1996 {Regression Shrinkage and Selection via the LASSO}.
\newblock \emph{J. of Roy. Statistical Soc.} \textbf{58}, 267--288.
\newblock ISSN 13697412.
\newblock (\doi{10.1111/j.1467-9868.2011.00771.x}).

\bibitem{Schmidt2009science}
Schmidt, M. \& Lipson, H., 2009 Distilling free-form natural laws from
  experimental data.
\newblock \emph{Science} \textbf{324}, 81--85.

\bibitem{Buchel2013}
B{\"{u}}chel, F., Rodriguez, N., Swainston, N., Wrzodek, C., Czauderna, T.,
  Keller, R., Mittag, F., Schubert, M., Glont, M., Golebiewski, M.
  \emph{et~al.}, 2013 {Path2Models: large-scale generation of computational
  models from biochemical pathway maps.}
\newblock \emph{BMC systems biology} \textbf{7}, 116.
\newblock ISSN 1752-0509.
\newblock (\doi{10.1186/1752-0509-7-116}).

\bibitem{Cohen2015}
Cohen, P.~R., 2015 {DARPA's Big Mechanism program.}
\newblock \emph{Physical biology} \textbf{12}, 045008.
\newblock ISSN 1478-3975.
\newblock (\doi{10.1088/1478-3975/12/4/045008}).

\bibitem{bishop2006pattern}
Bishop, C.~M. \& Others, 2006 \emph{{Pattern recognition and machine
  learning}}, volume~1.
\newblock Springer New York.

\bibitem{linde2005dic}
Linde, A., 2005 {DIC} in variable selection.
\newblock \emph{Statistica Neerlandica} \textbf{59}, 45--56.

\bibitem{rissanen1978modeling}
Rissanen, J., 1978 Modeling by shortest data description.
\newblock \emph{Automatica} \textbf{14}, 465--471.

\bibitem{Bongard2007pnas}
Bongard, J. \& Lipson, H., 2007 Automated reverse engineering of nonlinear
  dynamical systems.
\newblock \emph{Proceedings of the National Academy of Sciences} \textbf{104},
  9943--9948.

\bibitem{Quade2016arxiv}
Quade, M., Abel, M., Shafi, K., Niven, R.~K. \& Noack, B.~R., 2016 Prediction
  of dynamical systems by symbolic regression.
\newblock \emph{arXiv preprint arXiv:1602.04648} .

\bibitem{Duriez2016book}
Duriez, T., Brunton, S.~L. \& Noack, B.~R., 2016 \emph{Machine Learning
  Control: Taming Nonlinear Dynamics and Turbulence}.
\newblock Springer.

\bibitem{Wang2011prl}
Wang, W.~X., Yang, R., Lai, Y.~C., Kovanis, V. \& Grebogi, C., 2011 Predicting
  catastrophes in nonlinear dynamical systems by compressive sensing.
\newblock \emph{Physical Review Letters} \textbf{106}, 154101--1--154101--4.

\bibitem{Tran2016arxiv}
Tran, G. \& Ward, R., 2016 Exact recovery of chaotic systems from highly
  corrupted data.
\newblock \emph{arXiv preprint arXiv:1607.01067} .

\bibitem{Loiseau2016arxiv}
Loiseau, J.-C. \& Brunton, S.~L., 2016 Constrained sparse {Galerkin}
  regression.
\newblock \emph{arXiv preprint arXiv:1611.03271} .

\bibitem{Kutz2016book}
Kutz, J.~N., Brunton, S.~L., Brunton, B.~W. \& Proctor, J.~L., 2016
  \emph{Dynamic Mode Decomposition: Data-Driven Modeling of Complex Systems}.
\newblock SIAM.

\bibitem{burnham2004}
Burnham, K.~P. \& Anderson, R., 2004 {Multimodel Inference: Understanding AIC
  and BIC in Model Selection}.
\newblock \emph{Sociological Methods {\&} Research} \textbf{33}, 261--304.
\newblock ISSN 0049-1241.
\newblock (\doi{10.1177/0049124104268644}).

\bibitem{Lorenz1963}
Lorenz, E.~N. \& Lorenz, E.~N., 1963 {Deterministic Nonperiodic Flow}.
\newblock \emph{Journal of the Atmospheric Sciences} \textbf{20}, 130--141.
\newblock ISSN 0022-4928.
\newblock (\doi{10.1175/1520-0469(1963)020<0130:DNF>2.0.CO;2}).

\bibitem{gray}
Hey, T., Tansley, S., Tolle, K.~M. \emph{et~al.}, 2009 \emph{The fourth
  paradigm: data-intensive scientific discovery}, volume~1.
\newblock Microsoft research Redmond, WA.

\end{thebibliography}

\end{document}